\documentclass[conference]{IEEEtran}
\IEEEoverridecommandlockouts
\usepackage{amssymb}
\usepackage{amsmath}
\usepackage{amsthm}
\usepackage{amsfonts}
\usepackage{algorithm}
\usepackage{algorithmic}
\usepackage{array}
\usepackage{bbding}
\usepackage{bigints}
\usepackage{booktabs}
\usepackage{cite}
\usepackage{cleveref}
\usepackage{color}
\usepackage{diagbox}
\usepackage{epsfig,latexsym}
\usepackage{epstopdf}
\usepackage{graphicx}
\usepackage{fancyhdr}
\usepackage{float}
\usepackage{flushend}
\usepackage{indentfirst}
\usepackage{lastpage}
\usepackage{makecell}
\usepackage{mathtools}
\usepackage{multirow}
\usepackage{pifont}
\usepackage{psfrag}
\usepackage{setspace}
\usepackage{stfloats}
\usepackage{subfloat}
\usepackage{times}
\usepackage{arydshln}
\usepackage{textcomp}
\usepackage{url}
\usepackage{verbatim}
\usepackage[caption=false,font=normalsize,labelfont=sf,textfont=sf]{subfig}
\newcommand{\cmark}{\ding{52}}

\theoremstyle{remark}
\newtheorem{theorem}{\quad \textbf{Theorem}}

\allowdisplaybreaks[4]

\begin{document}
\title{Low-Complexity Channel Estimation for RIS-Assisted Multi-User Wireless Communications\\
\thanks{This work was supported by InterDigital. Lajos Hanzo would like to acknowledge the financial support of the Engineering and Physical Sciences Research Council (EPSRC) projects
under grant EP/Y037243/1, EP/W016605/1, EP/X01228X/1, EP/Y026721/1,
EP/W032635/1, EP/Y037243/1 and EP/X04047X/1 as well as of the European
Research Council's Advanced Fellow Grant QuantCom (Grant No. 789028).}}
\author{\IEEEauthorblockN{Qingchao Li\IEEEauthorrefmark{1},
Mohammed El-Hajjar\IEEEauthorrefmark{1},
Ibrahim Hemadeh\IEEEauthorrefmark{2},
Arman Shojaeifard\IEEEauthorrefmark{2},
and Lajos Hanzo\IEEEauthorrefmark{1}}
\IEEEauthorblockA{\IEEEauthorrefmark{1}Electronics and Computer Science,
University of Southampton, Southampton SO17 1BJ, U.K.\\
E-mail: qingchao.li@soton.ac.uk; meh@ecs.soton.ac.uk; lh@ecs.soton.ac.uk
}
\IEEEauthorblockA{\IEEEauthorrefmark{2}InterDigital, London EC2A 3QR, U.K.\\
E-mail: ibrahim.hemadeh@interdigital.com; arman.shojaeifard@interdigital.com
}
}
\maketitle

\begin{abstract}
Reconfigurable intelligent surfaces (RISs) are eminently suitable for improving the reliability of wireless communications by jointly designing the active beamforming at the base station (BS) and the passive beamforming at the RIS. Therefore, the accuracy of channel estimation is crucial for RIS-aided systems. The challenge is that only the cascaded two-hop channel spanning from the user equipments (UEs) to the RIS and spanning from the RIS to the BS can be estimated, due to the lack of active radio frequency (RF) chains at RIS elements, which leads to high pilot overhead. In this paper, we propose a low-overhead linear minimum mean square error (LMMSE) channel estimation method by exploiting the spatial correlation of channel links, which strikes a trade-off between the pilot overhead and the channel estimation accuracy. Moreover, we calculate the theoretical normalized mean square error (MSE) for our channel estimation method. Finally, we verify numerically that the proposed LMMSE estimator has lower MSE than the state-of-the-art (SoA) grouping based estimators.
\end{abstract}

\section{Introduction}
Reconfigurable intelligent surface (RIS) is a manmade surface comprised of a large number of reflecting elements for increasing the transmission reliability by jointly designing the active beamforming at the base station (BS) and the passive beamforming at the RIS~\cite{li2022enhancing}, \cite{li2022reconfigurable}, \cite{li2024ergodic}. To realize the advantage of RISs, accurate channel estimation is crucial~\cite{wang2023channel}, \cite{liu2021admm}, while this is challenging to achieve due to the lack of radio frequency (RF) chains at the RIS to process any pilot sequences. In the literature, a pair of popular channel estimation scenarios are advocated for RIS-aided systems, depending on the specific levels of channel state information (CSI) required, as discussed below.

In the first category, the joint BS and RIS beamforming designs are optimized based on the instantaneous CSI of the direct UE-BS and the cascaded UE-RIS-BS links~\cite{li2024low}, \cite{jensen2020optimal}, \cite{kundu2021channel}, \cite{li2023performance_tvt}, \cite{mutlu2024channel}. Li \textit{et al.}~\cite{li2024low} employed the linear mean square error (LMMSE) estimator for the instantaneous CSI acquisition, considering the hardware impairments (HWI) of the RIS configuration and the BS radio frequency (RF) chains. Jensen \textit{et al.}~\cite{jensen2020optimal} employed discrete Fourier transformation (DFT) matrix-based RIS training patterns for estimating the instantaneous channels having the classic least square (LS) algorithm. When the statistical channel information, such as the first and second moment of the channels, is available, the popular LMMSE estimation method can be employed for enhancing the channel estimation performance~\cite{kundu2021channel}, \cite{li2023performance_tvt}. In~\cite{mutlu2024channel}, Mutlu \textit{et al.} investigated the channel estimation for RIS-aided multi-user orthogonal frequency division multiplexing (OFDM) communication systems. Specifically, the cascaded channels are distinguished in the time domain using a reflection pattern that incorporates a discrete Fourier transform (DFT) matrix, while the users are separated in the frequency domain through the application of a Hadamard matrix.

However, again due to the lack of active signal processing units at the RIS, only the cascaded BS-RIS-UE channels can be estimated, where the required pilot sequence is proportional to the number of RIS elements. This means that the above instantaneous CSI estimation method has an excessive pilot overhead for a high number of RIS reflecting elements, which is impractical for short coherence intervals. To deal with this issue, the element grouping idea of~\cite{zheng2019intelligent}, \cite{zheng2020fast}, \cite{you2020channel}, \cite{yang2020intelligent} was adopted for the least square (LS) estimator to cut down pilot overhead by assigning the same phase to a group. However, the grouping concept is based on the assumption of having identical CSI for all the elements in the same group, hence resulting in obvious estimation accuracy degradation in practical RIS-aided scenarios.

To deal with the above problems, a novel low-complexity channel estimation method, termed as the correlated-grouping based LMMSE estimate, is proposed for RIS-aided systems. Explicitly, our contributions are summarized as follows. Firstly, we formulate the grouping-based low-overhead LMMSE estimator by exploiting the spatial correlation of the channel links. The proposed channel estimator can strike a flexible trade-off between the pilot overhead and the channel estimation accuracy. Secondly, the normalized mean square error (MSE) of the proposed correlated-grouping LMMSE estimator is derived theoretically. Finally, our numerical results show that the proposed correlated-grouping LMMSE estimator has better MSE performance than the state-of-the-art (SoA) grouping based estimators. Against this background, the novelty of the proposed low-overhead correlated-grouping LMMSE estimator is compared to the existing solutions in the literature~\cite{li2024low}, \cite{jensen2020optimal}, \cite{kundu2021channel}, \cite{li2023performance_tvt}, \cite{mutlu2024channel}, \cite{zheng2019intelligent}, \cite{zheng2020fast}, \cite{you2020channel}, \cite{yang2020intelligent} of the RIS-aided systems in Table~\ref{Table_literature}.

\begin{table*}[htb]
\small
\begin{center}
\caption{The novelty comparison of the proposed low-overhead correlated-grouping LMMSE estimator for the RIS-aided wireless communications to the existing solutions in the literatures~\cite{li2023performance_tvt}, \cite{jensen2020optimal}, \cite{kundu2021channel}, \cite{mutlu2024channel}, \cite{li2024low}, \cite{zheng2019intelligent}, \cite{zheng2020fast}, \cite{you2020channel}, \cite{yang2020intelligent}.}\label{Table_literature}
\begin{tabular}{*{12}{l}}
\hline
     & \makecell[c]{Our paper} & \makecell[c]{\cite{li2024low}} & \makecell[c]{\cite{jensen2020optimal}} & \makecell[c]{\cite{kundu2021channel}} & \makecell[c]{\cite{li2023performance_tvt}} & \makecell[c]{\cite{mutlu2024channel}} &  \makecell[c]{\cite{zheng2019intelligent}} & \makecell[c]{\cite{zheng2020fast}} & \makecell[c]{\cite{you2020channel}} & \makecell[c]{\cite{yang2020intelligent}} \\
\hline
    Multi-user & \makecell[c]{\cmark} & \makecell[c]{\cmark} & & & &  \makecell[c]{\cmark} &  &  & & \\
\hdashline
    Spatial correlation model for channel links & \makecell[c]{\cmark} & \makecell[c]{\cmark} & & & & & & & & \\
\hdashline
    Multiple BS antennas & \makecell[c]{\cmark} & \makecell[c]{\cmark} & \makecell[c]{\cmark} & \makecell[c]{\cmark} & \makecell[c]{\cmark} & \makecell[c]{\cmark} & &  &  &  \\
\hdashline
    Reduced pilot overhead & \makecell[c]{\cmark} &   & &  &  & & \makecell[c]{\cmark} & \makecell[c]{\cmark} & \makecell[c]{\cmark} & \makecell[c]{\cmark} \\
\hline
\end{tabular}
\end{center}
\end{table*}

\textit{Notations:} $(\cdot)^{\dag}$ and $(\cdot)^{\text{H}}$ represent the operation of conjugate and hermitian transpose, respectively, $\lceil a\rceil$ represents the minimal integer larger than $a$, $a_i$ and $a_{m,i}$ represents the $i$th element in vector $\mathbf{a}$ and $\mathbf{a}_{m}$, respectively, $[\mathbf{A}]_{n_1,n_2}$ represents the $(n_1,n_2)$th element in matrix $\mathbf{A}$, $\odot$ and $\otimes$ represent the Hadamard product and Kronecker product, respectively, $\mathbf{O}_{n_1\times n_2}$ represents the $(n_1\times n_2)$ zero matrix, $\mathbf{I}_{n}$ denotes the $n\times n$ identity matrix, $\text{Tr}[\mathbf{A}]$ represents the trace of $\mathbf{A}$, $\mathcal{CN}(\boldsymbol{\mu},\mathbf{\Sigma})$ denotes a circularly symmetric complex Gaussian random vector with mean $\boldsymbol{\mu}$ and covariance matrix $\mathbf{\Sigma}$, $\mathbb{E}[\mathbf{x}]$ represents the mean of a random vector $\mathbf{x}$, and $\mathbf{C}_{\mathbf{x}\mathbf{y}}$ represents the covariance matrix between the random vectors $\mathbf{x}$ and $\mathbf{y}$.

\section{System Model}\label{System_Model}
\begin{figure}[!t]
    \centering
    \includegraphics[width=3.3in]{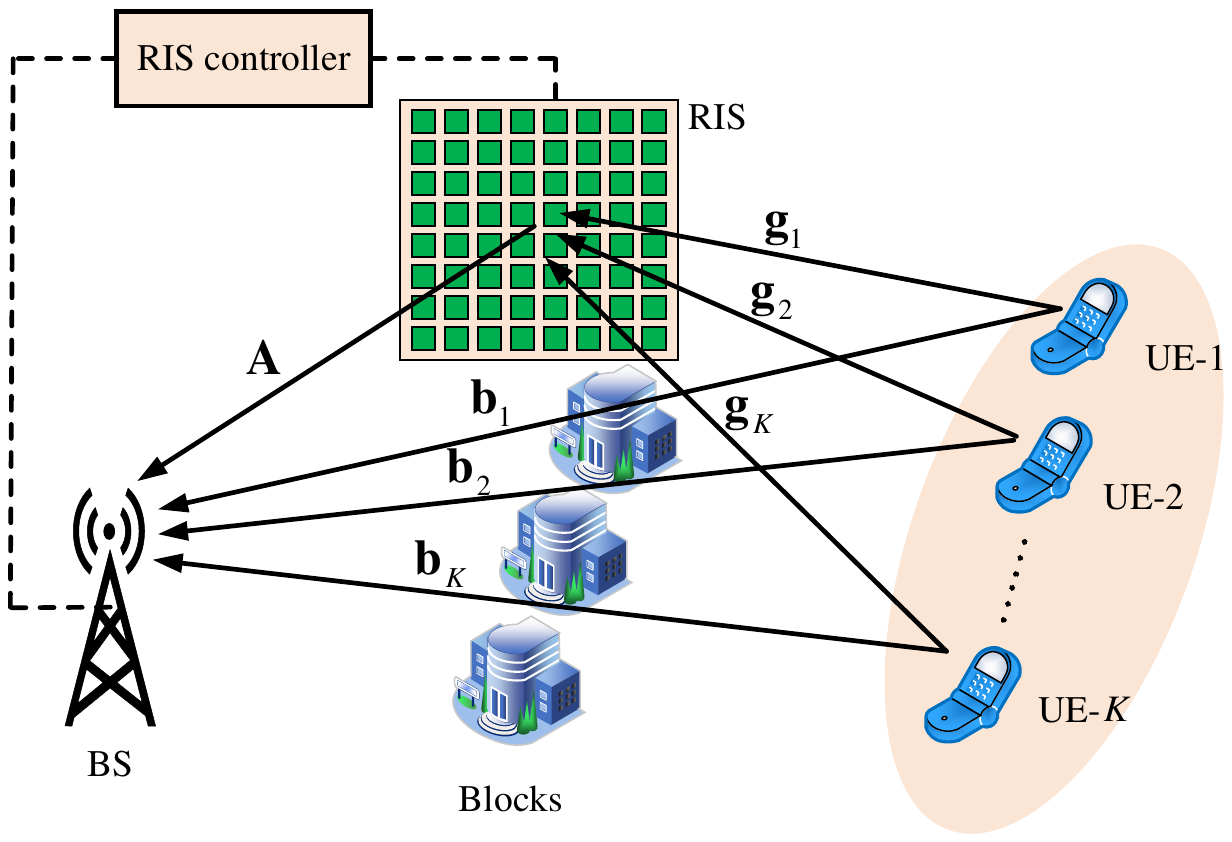}
    \caption{System model of the RIS-aided multi-user wireless communication system.}\label{System_two_stage_model}
\end{figure}

The model of our RIS-aided multi-user communication system is shown in Fig.~\ref{System_two_stage_model}, including a BS having $M$ linear array antennas, and $K$ single-antenna UEs, denoted as UE-1, UE-2, $\cdots$, UE-$K$. A uniform rectangular planar array (URPA) based RIS containing $N=N_x\times N_y$ reflecting elements, where $N_x$ and $N_y$ are the numbers of reflecting elements in the horizontal and the vertical direction, is used for increasing the transmission reliability. We denote the direct link from UE-$k$ to the BS as $\mathbf{b}_{k}\in\mathbb{C}^{M\times1}$, the link from UE-$k$ to the RIS as $\mathbf{g}_{k}\in\mathbb{C}^{N\times1}$, and the link from the RIS to the BS as $\mathbf{A}\in\mathbb{C}^{M\times N}$.

\subsection{RIS Architecture}
We denote the phase shift of the $n$th RIS element configured by the RIS controller as $\theta_n$, and the RIS phase shift vector can be represented as
\begin{align}
    \boldsymbol{\theta}=\left[\text{e}^{\jmath\theta_1},\text{e}^{\jmath\theta_2},\cdots,
    \text{e}^{\jmath\theta_N}\right].
\end{align}

\subsection{Channel Model}
We assume that the signals experience flat fading. The time-division
duplex (TDD) protocol is employed, and thus the channel responses are reciprocal. Since the RIS can be harnessed for creating additional paths for signal transmission among the BS and UEs, it is reasonable to assume that $\mathbf{g}_{k}$ and $\mathbf{A}$ experience Rician fading, while $\mathbf{b}_{k}$ is subject to Rayleigh fading. For the UE-BS links
\begin{align}
    \mathbf{b}_{k}\sim\mathcal{CN}\left(\mathbf{0}_M,\varrho_{\mathbf{b}_{k}}\mathbf{I}_M\right),
\end{align}
where $\varrho_{\mathbf{b}_{k}}$ is the large scale fading between UE-$k$ and the BS. For the UE-RIS links
\begin{align}
    \mathbf{g}_{k}\sim\mathcal{CN}\left(\sqrt{\frac{\kappa_{\mathbf{g}}\varrho_{\mathbf{g}_{k}}}
    {1+\kappa_{\mathbf{g}}}}\overline{\mathbf{g}}_{k},\frac{\varrho_{\mathbf{g}_{k}}}
    {1+\kappa_{\mathbf{g}}}\mathbf{R}_{k}\right),
\end{align}
where $\varrho_{\mathbf{g}_{k}}$ is the large scale fading between UE-$k$ and the RIS, $\kappa_{\mathbf{g}}$ is the Rician factor of the UE-RIS links, $\overline{\mathbf{g}}_{k}$ is the LoS component vector, each element of which has unit-modulus and its phase depends on the AoA of the path impinging on the RIS, denoted as $\varphi_{\text{a},k}$, and the distance of the $n_1$th and $n_2$th RIS elements, denoted as $\delta_{n_1,n_2}$. Furthermore, $\mathbf{R}_{k}$ is the covariance matrix of the non-line-of-sight (NLoS) component $\widetilde{\mathbf{g}}_{k}$. Since the RIS reflecting elements are tightly packed, $\widetilde{\mathbf{g}}_{k}$ is spatially correlated and $\mathbf{R}_{k}$ is not an identity matrix. For the RIS-BS links, we have
\begin{align}
    \mathbf{A}=\left[\mathbf{a}_1,\mathbf{a}_2,\cdots,\mathbf{a}_M\right]^\text{H},
\end{align}
where $\mathbf{a}_m\in\mathbb{C}^{N\times1}$ represents the channel response between the $m$th BS antenna and the RIS, given by
\begin{align}
    \mathbf{a}_m\sim\mathcal{CN}
    \left(\sqrt{\frac{\kappa_{\mathbf{A}}\varrho_{\mathbf{A}}}{1+\kappa_{\mathbf{A}}}}
    \overline{\mathbf{a}}_m,\frac{\varrho_{\mathbf{A}}}{1+\kappa_{\mathbf{A}}}\mathbf{R}_{0}\right),
\end{align}
where $\varrho_{\mathbf{A}}$ is the large scale fading between the BS and the RIS, $\kappa_{\mathbf{A}}$ is the Rician factor of the BS-RIS links, $\overline{\mathbf{a}}_m$ is the LoS component vector, each element of which has unit-modulus and its phase depends on the AoD from the RIS, denoted as $\varphi_\text{d}$, the AoA at the linear BS antenna array, denoted as $\psi$, and the distance of adjacent BS antennas, denoted as $\delta_0$. Furthermore, $\mathbf{R}_{0}$ is the covariance matrix of the NLoS component $\widetilde{\mathbf{a}}_m$.

\section{Correlated-grouping LMMSE estimator}
Since the TDD protocol is employed, each coherence interval is composed of $\tau_\text{c}$ symbol slots, which are assumed to have the same instantaneous CSI. To cut down the pilot overhead, we propose a novel scheme termed as correlated-grouping LMMSE estimator, which makes use of the statistical information of the correlated RIS-related channel links and it exhibits better MSE performance than the channel estimator based on the SoA grouping idea.

In this paper, we assume that the statistical CSI, which is $\overline{\mathbf{g}}_{k}$, $\overline{\mathbf{a}}_m$ and remain unchanged for a plenty of coherence intervals~\cite{wang2020compressed}, is known and our objective is to estimate the NLoS component in each coherence interval. A total of $T$ RIS training patterns, denoted as $\boldsymbol{\theta}_1,\boldsymbol{\theta}_2,\cdots,\boldsymbol{\theta}_T$, are activated in this order. Specifically, in each RIS pattern, $K$ symbol slots are assigned to the $K$ UEs transmitting carefully selected orthogonal pilot sequences to eliminate the inter-user interference, where we denote the pilot sequence transmitted in the $i$th symbol slot at the $k$th UE as $\phi_{k}^{(i)}$. Therefore, we have $\sum_{i=1}^{K}\phi_{k_1}^{(i)}\phi_{k_2}^{(i)\dag}=K$ when $k_1=k_2$ and $\sum_{i=1}^{K}\phi_{k_1}^{(i)}\phi_{k_2}^{(i)\dag}=0$ when $k_1\neq k_2$. In the $i$th symbol slot of the $t$th RIS pattern, the signal received at the BS is
\begin{align}
    \notag\mathbf{y}^{(t,i)}=&\sqrt{\rho_{k}}\sum_{k=1}^{K}
    \left[\sqrt{\varrho_{\mathbf{b}_{k}}}\mathbf{I}_{M},
    \sqrt{\varrho_{\mathbf{A}}\varrho_{\mathbf{g}_{k}}}\mathbf{I}_{M}
    \otimes\boldsymbol{\theta}_t\right]
    \mathbf{s}_{k}\phi_{k}^{(i)}\\
    &+\mathbf{w}^{(t,i)},
\end{align}
where $\rho_{k}$ is the transmitted pilot power of UE-$k$, $\mathbf{w}^{(t,i)}\sim\mathcal{CN}(\mathbf{0}_M,\sigma_w^2)$ is the additive white Gaussion noise (AWGN) at the BS, and
\begin{align}\label{Stage_I_1_2}
    \mathbf{s}_{k}=\left[\mathbf{b}_{k}^{\text{H}},\mathbf{r}_{k}^{(1)\text{H}},
    \mathbf{r}_{k}^{(2)\text{H}},\cdots,\mathbf{r}_{k}^{(M)\text{H}}\right]^{\text{H}},
\end{align}
in which $\mathbf{r}_{k}^{(m)}\in\mathbb{C}^{N\times1}$ is the cascaded channel given by
\begin{align}
    \mathbf{r}_{k}^{(m)}=\mathbf{a}_{m}\odot\mathbf{g}_{k}
    =[a_{m,1}g_{k,1},a_{m,2}g_{k,2},\cdots,a_{m,N}g_{k,N}]^\text{T}.
\end{align}
The UE-RIS channel $\mathbf{g}_{k}$ and the RIS-BS channel $\mathbf{A}$ cannot be estimated separately, since the RIS elements are passive, and our goal is to estimate $\mathbf{s}_{k}$. The received signal $\mathbf{y}^{(t,1)},\mathbf{y}^{(t,2)},
\cdots,\mathbf{y}^{(t,K)}$ can be combined with $\phi_k^{(1)\dag},\phi_k^{(2)\dag},\cdots,
\phi_k^{(K)\dag}$ respectively. Then we can get the observation for the channel related to UE-$k$ as
\begin{align}
    \mathbf{y}_{k}^{(t)}=\sum_{i=1}^{K}\mathbf{y}^{(t,i)}\phi_k^{(i)\dag}=\sqrt{\rho_{k}}
    \mathbf{Z}_k^{(t)}\mathbf{s}_{k}+\mathbf{w}_{k}^{(t)},
\end{align}
where
\begin{align}
    \mathbf{Z}_k^{(t)}=K\left[\sqrt{\varrho_{\mathbf{b}_{k}}}\mathbf{I}_{M},
    \sqrt{\varrho_{\mathbf{g}_{k}}\varrho_{\mathbf{A}}}\mathbf{I}_{M}
    \otimes\boldsymbol{\theta}_t\right],
\end{align}
and
\begin{align}
    \mathbf{w}_{k}^{(t)}=\sum_{i=1}^{K}\mathbf{w}^{(t,i)}\phi_k^{(i)\dag}\sim\mathcal{CN}
    \left(\mathbf{0}_M,K\sigma_w^2\right).
\end{align}
By stacking the observations $\mathbf{y}_{k}^{(1)},\mathbf{y}_{k}^{(2)},\cdots,\mathbf{y}_{k}^{(T)}$, we get
\begin{align}
    \mathbf{y}_{k}=\sqrt{\rho_{k}}\mathbf{Z}_{k}\mathbf{s}_{k}+\mathbf{w}_{k},
\end{align}
where the observation for UE-$k$ is
\begin{align}
    \mathbf{y}_{k}=\left[\mathbf{y}_{k}^{(1)\text{H}},\mathbf{y}_{k}^{(2)\text{H}},\cdots,
    \mathbf{y}_{k}^{(T)\text{H}}\right]^\text{H},
\end{align}
the noise in the observation for UE-$k$ is
\begin{align}
    \mathbf{w}_{k}=\left[\mathbf{w}_{k}^{(1)\text{H}},\mathbf{w}_{k}^{(2)\text{H}},\cdots,
    \mathbf{w}_{k}^{(T)\text{H}}\right]^\text{H}\!\sim\!\mathcal{CN}
    \left(\mathbf{0}_{MT},K\sigma_w^2\mathbf{I}_{MT}\right),
\end{align}
and
\begin{align}
    \mathbf{Z}_{k}=\left[\mathbf{Z}_k^{(1)\text{H}},\mathbf{Z}_k^{(2)\text{H}},\cdots,
    \mathbf{Z}_k^{(T)\text{H}}\right]^{\text{H}}.
\end{align}
After some further manipulations, the mean and covariance matrix of $\mathbf{s}_{k}$ are
\begin{align}
    \notag\mathbb{E}[\mathbf{s}_{k}]=
    &\sqrt{\frac{\kappa_{\mathbf{A}}\kappa_{\mathbf{g}_{k}}}
    {\left(1+\kappa_{\mathbf{A}}\right)\left(1+\kappa_{\mathbf{g}_{k}}\right)}}
    \left[\mathbf{0}_M^\mathrm{H},
    \left(\overline{\mathbf{a}}_1\odot\overline{\mathbf{g}}_{k}\right)^\mathrm{H},\cdots,\right.\\
    &\left.\left(\overline{\mathbf{a}}_{M}\odot
    \overline{\mathbf{g}}_{k}\right)^\mathrm{H}\right]^\mathrm{H}
\end{align}
and
\begin{align}\label{Stage_I_7}
    \notag&\mathbf{C}_{\mathbf{s}_{k}\mathbf{s}_{k}}
    \!\!=\!\!\left[\begin{array}{cccc}
             \mathbf{O}_{{M}\times {M}} & \mathbf{O}_{{M}\times {N}} & \cdots & \mathbf{O}_{{M}\times {N}}\\
             \mathbf{O}_{{N}\times {M}} & \overline{\mathbf{A}}_{1,1}\odot\mathbf{R}_{\mathbf{g}}^{(k)} & \cdots & \overline{\mathbf{A}}_{1,M}\odot\mathbf{R}_{\mathbf{g}}^{(k)} \\
             \vdots & \vdots & \ddots & \vdots\\
             \mathbf{O}_{{N}\times {M}} & \overline{\mathbf{A}}_{M,1}\odot\mathbf{R}_{\mathbf{g}}^{(k)} & \cdots & \overline{\mathbf{A}}_{M,M}\odot\mathbf{R}_{\mathbf{g}}^{(k)} \\
         \end{array}\right]\\
    &+\left[\begin{array}{cc}
         \mathbf{I}_{{M}} & \mathbf{O}_{{M}\times{MN}}\\
         \mathbf{O}_{{MN}\times {M}} & \mathbf{I}_{{M}}\otimes\left(\mathbf{R}_{\mathbf{A}}^{(0)}
         \odot\left(\overline{\mathbf{G}}^{(k)}+\mathbf{R}_{\mathbf{g}}^{(k)}\right)\right)
         \end{array}\right],
\end{align}
respectively, where $\mathbf{R}_{\mathbf{A}}^{(0)}=\frac{1}{1+\kappa_{\mathbf{A}}}\mathbf{R}_{0}$, $\mathbf{R}_{\mathbf{g}}^{(k)}=\frac{1}{1+\kappa_{\mathbf{g}}}\mathbf{R}_{k}$, $\overline{\mathbf{G}}^{(k)}=\frac{\kappa_{\mathbf{g}}}{1+\kappa_{\mathbf{g}}}
\overline{\mathbf{g}}_{k}\overline{\mathbf{g}}_{k}^{\text{H}}$ and $\overline{\mathbf{A}}_{m_1,m_2}=\frac{\kappa_{\mathbf{A}}}
{1+\kappa_{\mathbf{A}}}\overline{\mathbf{a}}_{m_1}\overline{\mathbf{a}}_{m_2}^{\text{H}}$.

The conventional LMMSE estimate of $\mathbf{s}_{k}$ is given by
\begin{align}\label{Stage_I_8}
    \hat{\mathbf{s}}_{k}=\mathbb{E}[\mathbf{s}_{k}]
    +\mathbf{C}_{\mathbf{s}_{k}\mathbf{y}_{k}}\mathbf{C}_{\mathbf{y}_{k}\mathbf{y}_{k}}^{-1}
    \left(\mathbf{y}_{k}-\mathbb{E}[\mathbf{y}_{k}]\right),
\end{align}
where
\begin{align}
    \mathbb{E}[\mathbf{y}_{k}]=\sqrt{\rho_{k}}\mathbf{Z}_{k}\mathbb{E}[\mathbf{s}_{k}],
\end{align}
\begin{align}
    \mathbf{C}_{\mathbf{s}_{k}\mathbf{y}_{k}}=\sqrt{\rho_{k}}\mathbf{C}_{\mathbf{s}_{k}
    \mathbf{s}_{k}}\mathbf{Z}_{k}^{\mathrm{H}}
\end{align}
and
\begin{align}
    \mathbf{C}_{\mathbf{y}_{k}\mathbf{y}_{k}}
    =\rho_{k}\mathbf{Z}_{k}\mathbf{C}_{\mathbf{s}_{k}\mathbf{s}_{k}}\mathbf{Z}_{k}^\mathrm{H}
    +K\sigma_{w}^2\mathbf{I}_{MT}.
\end{align}
Since $\mathbf{s}_{k}$ is an $M(N+1)\times1$ vector and $\mathbf{Z}_{k}$ is an $MT\times M(N+1)$ matrix, the number of RIS training patterns must satisfy that $T\geq N+1$ for uniquely estimating $\mathbf{s}_{k}$. Thus, at least $\tau_\text{p}=K(N+1)$ symbol slots are required for instantaneous CSI estimation. For the RIS elements having discrete phase shifts, the orthogonal RIS training patterns $\boldsymbol{\theta}_{1},\boldsymbol{\theta}_{2},\cdots$,$\boldsymbol{\theta}_{T}$ can be designed based on a Hadamard matrix. Specifically, $[1,\boldsymbol{\theta}_{t}]$ represents the first $N+1$ elements in the $t$th row of the $2^{\lceil\log_2T\rceil}\times2^{\lceil\log_2T\rceil}$ Hadamard matrix.

To cut down the pilot overhead, the RIS grouping idea was developed in~\cite{zheng2019intelligent}, \cite{zheng2020fast}, \cite{you2020channel}, \cite{yang2020intelligent}, in which the $N$ RIS elements are divided into $N_{\mathcal{G}}$ groups. It was assumed that the channel links corresponding to the reflecting elements in the same group have identical CSI. Thus, the number of RIS training patterns just satisfy that $T\geq N_\mathcal{G}+1$. However, this grouping method is based on the assumption that the channel response corresponding to the RIS elements in the same group are identical, and that in the different groups are independent. In practice, this assumption is difficult to satisfy, because this grouping method is based on the assumption that the spatial correlation matrix obeys
\begin{align}
    \mathbf{R}_{k}=\mathbf{I}_{N_\mathcal{G}}
    \otimes\left(\mathbf{1}_{\frac{N}{N_\mathcal{G}}}
    \cdot\mathbf{1}_{\frac{N}{N_\mathcal{G}}}^{\text{H}}\right),
\end{align}
where $\mathbf{1}_{\frac{N}{N_\mathcal{G}}}$ is the $(\frac{N}{N_\mathcal{G}}\times1)$ vector having all elements of 1. This is the ideal spatial correlation model, but the exponential correlation model is more suitable in practice~\cite{hampton2013introduction}, \cite{bithas2014exploiting}. Specifically, in the exponential correlation model the $(n_1,n_2)$th element in $\mathbf{R}_{k}$ is
\begin{align}
    [\mathbf{R}_{k}]_{n_1,n_2}=\eta_k^{\frac{\delta_{n_1,n_2}}{\lambda}},
\end{align}
where $\eta_k$ ($0\leq\eta_k\leq1$) is the correlation coefficient of $\widetilde{\mathbf{g}}_{k}$ (if $k=1,2,\cdots,K$) or $\widetilde{\mathbf{a}}_m$ (if $k=0$) at the reference distance of the wavelength $\lambda$. To make the Grouping LMMSE method work for the practical correlation model, we propose a novel low-overhead channel estimation scheme, namely the correlated-grouping LMMSE estimator, detailed as follows.

Firstly, we partition the $N$ RIS elements into $N_\mathcal{G}$ groups, each of which contains $\frac{N}{N_\mathcal{G}}$ elements. In the RIS training patterns, if the RIS elements in the same group have identical phase shift, the sum of NLoS CSIs in the same group can be estimated by the conventional LMMSE method, with the number of RIS training patterns satisfying $T\geq N_\mathcal{G}+1$. Specifically, we denote index set of RIS elements in the $g$th group as $\mathcal{I}_g$. For simplicity, we denote the indices of RIS elements that are sorted according to the grouping division, i.e. $\mathcal{I}_1=\{1,2,\cdots,\frac{N}{N_\mathcal{G}}\}$, $\mathcal{I}_2=\{\frac{N}{N_\mathcal{G}}+1,\frac{N}{N_\mathcal{G}}+2,\cdots,
2\frac{N}{N_\mathcal{G}}\}$ and up to $\mathcal{I}_{N_\mathcal{G}}
=\{(N_\mathcal{G}-1)\frac{N}{N_\mathcal{G}}+1,(N_\mathcal{G}-1)\frac{N}{N_\mathcal{G}}+2,
\cdots,N\}$. The $t$th RIS training pattern is given by
\begin{align}
    \boldsymbol{\theta}_{t}
    =\boldsymbol{\theta}_{\mathcal{G},t}\otimes\mathbf{1}_{\frac{N}{N_\mathcal{G}}}^{\text{H}},
\end{align}
where $[1,\boldsymbol{\theta}_{\mathcal{G},t}]$ represents the first $N_{\mathcal{G}}+1$ elements in the $t$th row of the $2^{\lceil\log_2T\rceil}\times2^{\lceil\log_2T\rceil}$ Hadamard matrix. We define
\begin{align}
    \mathbf{u}_{k}=\left[\mathbf{b}_{k}^{\text{H}},\mathbf{v}_k^{(1)\text{H}},
    \mathbf{v}_k^{(2)\text{H}},\cdots,\mathbf{v}_{k}^{(M)\text{H}}\right]^\text{H},
\end{align}
where $v_{k,g}^{(m)}=\sum_{l\in\mathcal{I}_g}(r_{k,l}^{(m)}-\mathbb{E}[r_{k,l}^{(m)}])$, with $g=1,2,\cdots,N_{\mathcal{G}}$, respectively. The relationship between $\mathbf{y}_{k}$ and $\mathbf{u}_{k}$ is
\begin{align}
    \mathbf{y}_{k}=\sqrt{\rho_{k}}\mathbf{Z}_{\mathcal{G},k}\mathbf{u}_{k}+\mathbf{w}_{k},
\end{align}
where
\begin{align}
    \mathbf{Z}_{\mathcal{G},k}=\left[\mathbf{Z}_{\mathcal{G},k}^{(1)\text{H}},
    \mathbf{Z}_{\mathcal{G},k}^{(2)\text{H}},\cdots,\mathbf{Z}_{\mathcal{G},k}^{(T)\text{H}}\right]
    ^{\text{H}}
\end{align}
with
\begin{align}
    \mathbf{Z}_{\mathcal{G},k}^{(t)}=K\left[\sqrt{\varrho_{\mathbf{b}_{k}}}\mathbf{I}_{M},
    \sqrt{\varrho_{\mathbf{g}_{k}}\varrho_{\mathbf{A}}}\mathbf{I}_{M}\otimes
    \boldsymbol{\theta}_{\mathcal{G},t}\right].
\end{align}
The mean of $\mathbf{u}_{k}$ is $\mathbb{E}[\mathbf{u}_{k}]=\mathbf{0}_{M(N_\mathcal{G}+1)\times1}$, and its covariance matrix $\mathbf{C}_{\mathbf{u}_{k}\mathbf{u}_{k}}$ can be directly determined from $\mathbf{C}_{\mathbf{s}_{k}\mathbf{s}_{k}}$ by summing up the elements with the row indices and column indices corresponding to the same group. Therefore, the LMMSE estimate of $\mathbf{u}_{k}$ is
\begin{align}\label{Stage_I_15}
    \hat{\mathbf{u}}_{k}=\mathbf{C}_{\mathbf{u}_{k}\mathbf{y}_{k}}\mathbf{C}_{\mathbf{y}_{k}
    \mathbf{y}_{k}}^{-1}\left(\mathbf{y}_{k}-\mathbb{E}[\mathbf{y}_{k}]\right),
\end{align}
where $\mathbf{C}_{\mathbf{u}_{k}\mathbf{y}_{k}}
=\sqrt{\rho_{k}}\mathbf{C}_{\mathbf{u}_{k}\mathbf{u}_{k}}\mathbf{Z}_{\mathcal{G},k}^{\mathrm{H}}$.

Secondly, based on $\hat{\mathbf{u}}_{k}$, the LMMSE estimation of $\mathbf{s}_{k}$ is
\begin{align}
    \hat{\mathbf{s}}_{\mathcal{G},k}=\mathbf{C}_{\mathbf{s}_{k}\hat{\mathbf{u}}_{k}}
    \mathbf{C}_{\hat{\mathbf{u}}_{k}\hat{\mathbf{u}}_{k}}^{-1}
    \left(\hat{\mathbf{u}}_{k}-\mathbb{E}[\hat{\mathbf{u}}_{k}]\right),
\end{align}
where $\mathbb{E}[\hat{\mathbf{u}}_{k}]=\mathbf{0}_{M(N_\mathcal{G}+1)\times1}$. Then, based on (\ref{Stage_I_15}) we can get
\begin{align}
    \mathbf{C}_{\hat{\mathbf{u}}_{k}\hat{\mathbf{u}}_{k}}=\mathbf{C}_{\mathbf{u}_{k}\mathbf{y}_{k}}
    \mathbf{C}_{\mathbf{y}_{k}\mathbf{y}_{k}}^{-1}\mathbf{C}_{\mathbf{u}_{k}\mathbf{y}_{k}}^{\text{H}}
\end{align}
and
\begin{align}
    \mathbf{C}_{\mathbf{s}_{k}\hat{\mathbf{u}}_{k}}=\mathbf{C}_{\mathbf{s}_{k}\mathbf{y}_{k}}
    \mathbf{C}_{\mathbf{y}_{k}\mathbf{y}_{k}}^{-1}\mathbf{C}_{\mathbf{u}_{k}\mathbf{y}_{k}}
    ^{\text{H}}.
\end{align}
Therefore, we can get
\begin{align}
    \notag\hat{\mathbf{s}}_{\mathcal{G},k}
    =&\mathbb{E}[\mathbf{s}_{k}]+\mathbf{C}_{\mathbf{s}_{k}\hat{\mathbf{u}}_{k}}
    \mathbf{C}_{\hat{\mathbf{u}}_{k}\hat{\mathbf{u}}_{k}}^{-1}
    \hat{\mathbf{u}}_{k}\\
    \notag=&\mathbb{E}[\mathbf{s}_{k}]+\mathbf{C}_{\mathbf{s}_{k}
    \mathbf{y}_{k}}\mathbf{C}_{\mathbf{y}_{k}\mathbf{y}_{k}}^{-1}\mathbf{C}_{\mathbf{u}_{k}
    \mathbf{y}_{k}}^{\text{H}}\left(\mathbf{C}_{\mathbf{u}_{k}\mathbf{y}_{k}}
    \mathbf{C}_{\mathbf{y}_{k}\mathbf{y}_{k}}^{-1}\cdot\right.\\
    &\left.\mathbf{C}_{\mathbf{u}_{k}\mathbf{y}_{k}}^{\text{H}}\right)^{-1}
    \mathbf{C}_{\mathbf{u}_{k}\mathbf{y}_{k}}
    \mathbf{C}_{\mathbf{y}_{k}\mathbf{y}_{k}}^{-1}
    \left(\mathbf{y}_{k}-\mathbb{E}[\mathbf{y}_{k}]\right).
\end{align}
Upon define the estimation error of $\mathbf{s}_{k}$ as $\widetilde{\mathbf{s}}_{k}=\mathbf{s}_{k}-\hat{\mathbf{s}}_{\mathcal{G},k}$, the estimation error covariance matrix can be derived as
\begin{align}\label{Stage_I_21}
    \notag\mathbf{C}_{\widetilde{\mathbf{s}}_{k}\widetilde{\mathbf{s}}_{k}}
    =&\mathbf{C}_{\mathbf{s}_{k}\mathbf{s}_{k}}-\mathbf{C}_{\mathbf{s}_{k}
    \mathbf{y}_{k}}\mathbf{C}_{\mathbf{y}_{k}\mathbf{y}_{k}}^{-1}
    \mathbf{C}_{\mathbf{u}_{k}\mathbf{y}_{k}}^{\text{H}}\cdot\\
    &\left(\mathbf{C}_{\mathbf{u}_{k}\mathbf{y}_{k}}
    \mathbf{C}_{\mathbf{y}_{k}\mathbf{y}_{k}}^{-1}\mathbf{C}_{\mathbf{u}_{k}
    \mathbf{y}_{k}}^{\text{H}}\right)^{-1}
    \mathbf{C}_{\mathbf{u}_{k}\mathbf{y}_{k}}\mathbf{C}_{\mathbf{y}_{k}
    \mathbf{y}_{k}}^{-1}\mathbf{C}_{\mathbf{s}_{k}\mathbf{y}_{k}}^{\text{H}},
\end{align}
and the normalized MSE of the estimated $\mathbf{s}_{k}$, denoted as $\epsilon_k$, is
\begin{align}
    \notag\epsilon_k=&\text{Tr}\left[\left(\mathbf{s}_{k}-\hat{\mathbf{s}}_{\mathcal{G},k}\right)
    \left(\mathbf{s}_{k}-\hat{\mathbf{s}}_{\mathcal{G},k}\right)^{\text{H}}\right]\\
    &=\text{Tr}\left[\mathbf{C}_{\widetilde{\mathbf{s}}_{k}\widetilde{\mathbf{s}}_{k}}\right].
\end{align}

\begin{theorem}\label{Theorem_1}
When the transmit power obeys $\rho_k\rightarrow\infty$, the normalized MSE of the estimated $\mathbf{s}_{k}$, denoted as $\epsilon_k^{(\rho_k\rightarrow\infty)}$, is given by
\begin{align}\label{Performance_Analysis_2}
    \notag &\epsilon_k^{(\rho_k\rightarrow\infty)}=
    \text{Tr}\left[\mathbf{C}_{\mathbf{s}_{k}\mathbf{s}_{k}}
    -\mathbf{C}_{\mathbf{s}_{k}\mathbf{s}_{k}}\mathbf{Z}_{k}^{\mathrm{H}}
    \left(\mathbf{Z}_{k}\mathbf{C}_{\mathbf{s}_{k}\mathbf{s}_{k}}
    \mathbf{Z}_{k}^{\mathrm{H}}\right)^{-1}
    \mathbf{Z}_{\mathcal{G},k}\cdot\right.\\
    \notag&\mathbf{C}_{\mathbf{u}_{k}\mathbf{u}_{k}}^{\text{H}}
    \left.\left(\mathbf{C}_{\mathbf{u}_{k}\mathbf{u}_{k}}
    \mathbf{Z}_{\mathcal{G},k}^{\mathrm{H}}
    \left(\mathbf{Z}_{k}\mathbf{C}_{\mathbf{s}_{k}\mathbf{s}_{k}}
    \mathbf{Z}_{k}^{\mathrm{H}}\right)^{-1}\mathbf{Z}_{\mathcal{G},k}
    \mathbf{C}_{\mathbf{u}_{k}\mathbf{u}_{k}}^{\text{H}}\right)^{-1}\cdot\right.\\
    &\left.\mathbf{C}_{\mathbf{u}_{k}\mathbf{u}_{k}}\mathbf{Z}_{\mathcal{G},k}^{\mathrm{H}}
    \left(\mathbf{Z}_{k}\mathbf{C}_{\mathbf{s}_{k}\mathbf{s}_{k}}
    \mathbf{Z}_{k}^{\mathrm{H}}\right)^{-1}
    \mathbf{Z}_{k}\mathbf{C}_{\mathbf{s}_{k}\mathbf{s}_{k}}^{\mathrm{H}}\right].
\end{align}
\end{theorem}
\begin{IEEEproof}
    It can be directly obtained by setting $\rho_k\rightarrow\infty$ for $\mathbf{C}_{\mathbf{s}_{k}\mathbf{y}_{k}}=\sqrt{\rho_{k}}\mathbf{C}_{\mathbf{s}_{k}
    \mathbf{s}_{k}}\mathbf{Z}_{k}^{\mathrm{H}}$, $\mathbf{C}_{\mathbf{u}_{k}\mathbf{y}_{k}}
    =\sqrt{\rho_{k}}\mathbf{C}_{\mathbf{u}_{k}\mathbf{u}_{k}}
    \mathbf{Z}_{\mathcal{G},k}^{\mathrm{H}}$ and $\mathbf{C}_{\mathbf{y}_{k}\mathbf{y}_{k}}
    =\rho_{k}\mathbf{Z}_{k}\mathbf{C}_{\mathbf{s}_{k}\mathbf{s}_{k}}\mathbf{Z}_{k}^\mathrm{H}
    +K\sigma_{w}^2\mathbf{I}_{MT}$ in (\ref{Stage_I_21}).
\end{IEEEproof}

\section{Simulation Results}\label{Simulation_Results}
In our system, the BS and the RIS are located at the Cartesian coordinates of (0m, 0m, 15m) and (0m, 50m, 10m), respectively. Furthermore, $K=4$ UEs are deployed at the Cartesian coordinates of (-8m, 44m, 5m), (-6m, 42m, 5m), (6m, 42m, 5m) and (8m, 44m, 5m). The path loss is described by distance-dependent model, given by $\varrho=\varrho_0d^{-\alpha}$, where $\varrho_0$ is the path loss at the reference distance of 1 meter, $d$ and $\alpha$ denote the distance between nodes (BS, RIS and UEs) and the corresponding path loss exponent. In order to facilitate the comparison of the performance in various scenarios for RIS-aided wireless communications, we assume that the direct UE-BS link is completely blocked for characterizing the performance benefits of the RIS. Referring to~\cite{you2020channel}, the simulation parameters are: $\kappa_\mathbf{A}=-20\text{dB}$, $\kappa_\mathbf{g}=3\text{dB}$, $\alpha_\mathbf{A}=2.5$, $\alpha_\mathbf{g}=2.2$, $\varrho_0=-30\text{dB}$, $\sigma_w^2=-89\text{dBm}$, $\delta_x=\delta_y=\frac{\lambda}{2}$, $\delta_0=\frac{\lambda}{2}$, unless otherwise specified. Furthermore, we have $M=8$, $N=8\times8$, $\psi=\frac{\pi}{3}$, $\rho_{1}=\rho_{2}=\cdots=\rho_{K}=\rho$, $\eta_0=\eta_1=\cdots=\eta_K=0.99$, the values of $\varphi_{\text{a},1},\varphi_{\text{a},2},\cdots,\varphi_{\text{a},K}$ and $\varphi_\text{d}$ are calculated based on the position of the UEs, RIS and BS.

\begin{figure}[!t]
    \centering
    \includegraphics[width=3.2in]{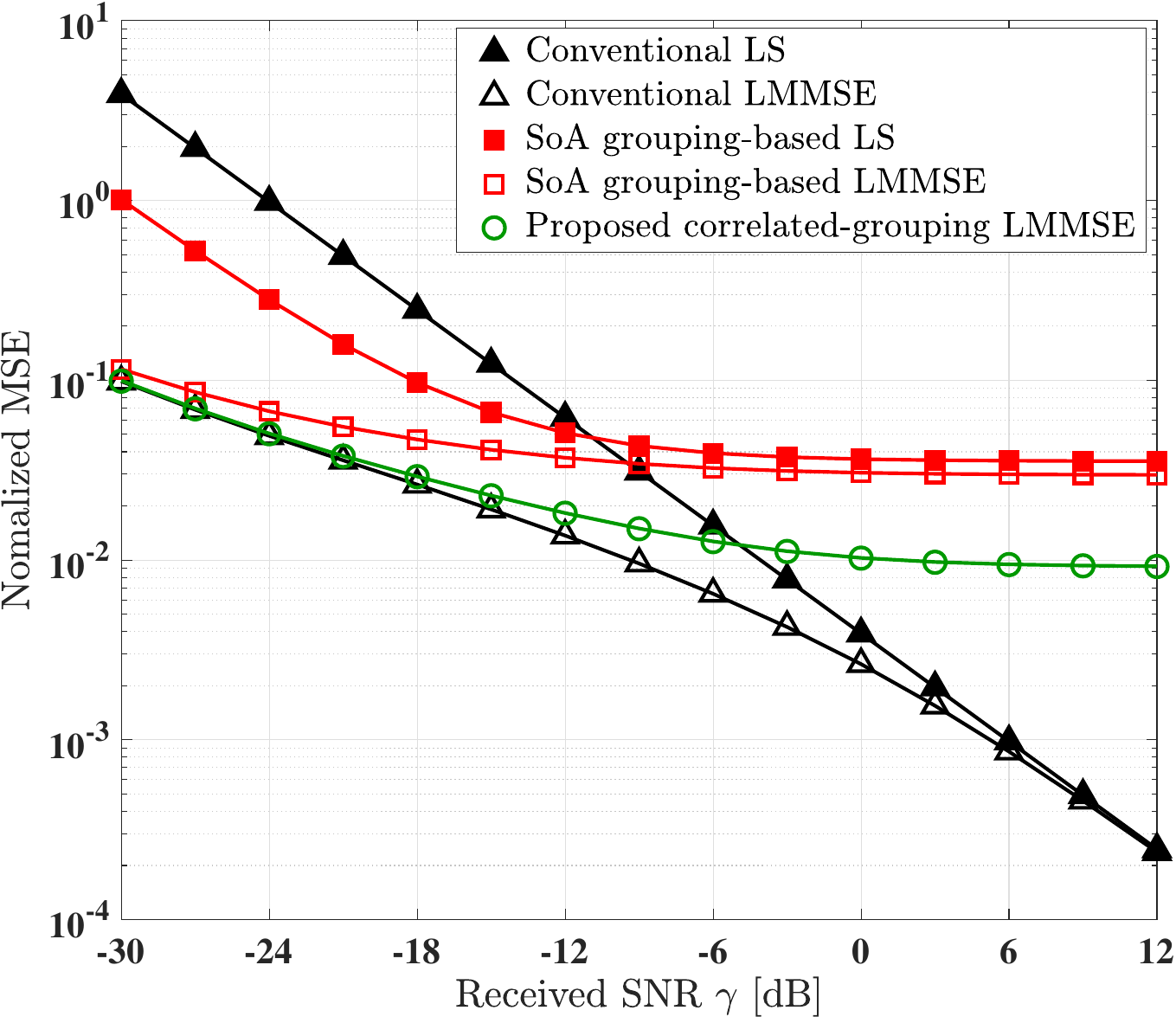}
    \caption{Theoretical analysis and simulation results of the received SNR $\gamma$ versus the normalized MSE in different estimators, where the solid lines represent the theoretical results.}\label{Simu_fig_1}
\end{figure}

\begin{figure}[!t]
    \centering
    \includegraphics[width=3.2in]{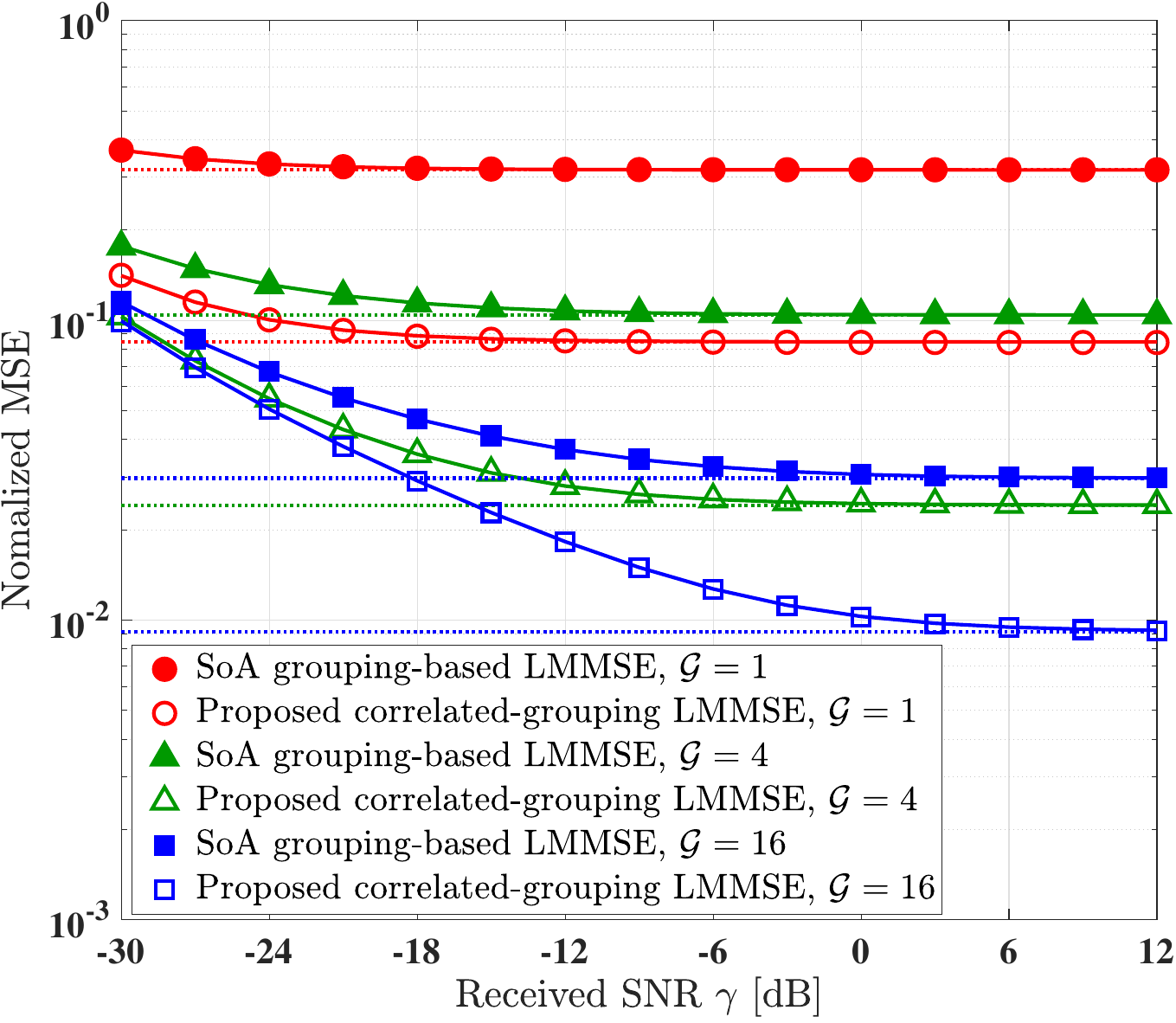}
    \caption{Theoretical analysis and simulation results of the received SNR $\gamma$ versus the normalized MSE with different number of groups $\mathcal{G}$, where the solid lines represent the theoretical results and the the solid lines represent the theoretical bound with the transmit power $\rho_k\rightarrow\infty$.}\label{Simu_fig_2}
\end{figure}

Firstly, the theoretical analysis and simulation results of the normalized MSE of the different estimators versus the received signal-to-noise ratio (SNR) $\gamma$ are characterized in Fig.~\ref{Simu_fig_1}. These include the conventional LS/LMMSE estimators in~\cite{jensen2020optimal}, \cite{kundu2021channel}, \cite{li2023performance_tvt}, the SoA grouping-based LS/LMMSE estimators in~\cite{zheng2019intelligent}, \cite{zheng2020fast}, \cite{you2020channel}, \cite{yang2020intelligent}, and our proposed correlated-grouping LMMSE estimators, where the number of groups $\mathcal{G}=16$. Fig.~\ref{Simu_fig_1} shows that our proposed correlated-grouping LMMSE outperforms the SoA grouping-based LMMSE, since the realistic exponential spatial correlation model is considered in our proposed scheme. Furthermore, observe in the low transmit power region of Fig.~\ref{Simu_fig_1} that the MSE performance of our proposed correlated-grouping LMMSE estimators is comparable to that of the conventional LMMSE estimator. However, in the high transmit power region, the normalized MSE of the grouping-based LMMSE estimators tend to a floor due to the reduced number of RIS training patterns.

Fig.~\ref{Simu_fig_2} compares the MSE performance of the SoA grouping-based LMMSE and our proposed correlated-grouping LMMSE estimators for different number of groups $\mathcal{G}$. Explicitly, it shows that our proposed correlated-grouping LMMSE estimator has a  significantly lower MSE than that of the SoA grouping-based LMMSE. This is due to the fact that the SoA grouping-based LMMSE method assumes that the NLoS CSI corresponding to the RIS elements in the same group is identical, while the statistical information in our proposed correlated-grouping LMMSE method is based on the realistic spatial correlation of RIS elements.

\section{Conclusions}\label{Conclusions}
In this paper, we addressed the problem of high-overhead channel estimation in RIS-aided multi-user systems. To this end, we formulated a novel low-complexity estimator, namely the correlated-grouping LMMSE estimator, by exploiting the spatial correlation of the RIS channel links. Numerical results showed the our proposed correlated-grouping LMMSE estimator has lower MSE than the SoA grouping based LS estimators.

\bibliographystyle{IEEEtran}
\bibliography{IEEEabrv,TAMS}
\end{document}